\documentclass[conference,9pt]{IEEEtran}

\AtBeginDocument{%
  \providecommand\BibTeX{{%
    \normalfont B\kern-0.5em{\scshape i\kern-0.25em b}\kern-0.8em\TeX}}}

\usepackage{
  latexsym,        
  color,           
  xcolor,          
  multirow,        
  array,           
  listings,        
  enumitem,        
  multicol,        
  balance,         
  graphicx,        
  hyperref,        
  url,             
  algorithmic,     
  textcomp         
}
\usepackage[normalem]{ulem}
\usepackage{subcaption}
\usepackage[linesnumbered,ruled,vlined]{algorithm2e}
\SetKw{Continue}{continue}  
\usepackage{tikz}
\usepackage{amsmath}
\usetikzlibrary{positioning}

\newcolumntype{P}[1]{>{\centering\arraybackslash}p{#1}}

\newtheorem{auxdefn}{Definition}[section]

\newtheorem{auxexample}{Example}[section]

\begin{document}

\title{An LLVM-Based Optimization Pipeline for SPDZ}




\author{
    \IEEEauthorblockN{Tianye Dai}
    \IEEEauthorblockA{
        \textit{Davidson College}\\
        Davidson, NC, USA \\
        asdai@davidson.edu
    }
    \and
    \IEEEauthorblockN{Hammurabi Mendes}
    \IEEEauthorblockA{
        \textit{Davidson College}\\
        Davidson, NC, USA \\
        hamendes@davidson.edu
    }
    \and
    \IEEEauthorblockN{HeuiChan Lim}
    \IEEEauthorblockA{
        \textit{Davidson College}\\
        Davidson, NC, USA \\
        telim@davidson.edu
    }
}


\maketitle

\begin{abstract}
Actively secure arithmetic MPC is now practical for real applications, but performance and usability are still limited by framework-specific compilation stacks, the need for programmers to explicitly express parallelism, and high communication overhead. We design and implement a proof-of-concept LLVM-based optimization pipeline for the SPDZ protocol that addresses these bottlenecks. Our front end accepts a subset of C with lightweight privacy annotations and lowers it to LLVM IR, allowing us to reuse mature analyses and transformations to automatically batch independent arithmetic operations. Our back end performs data-flow and control-flow analysis on the optimized IR to drive a non-blocking runtime scheduler that overlaps independent operations and aggressively overlaps communication with computation; when enabled, it can map batched operations to GPU kernels. This design preserves a low learning curve by using a mainstream language and hiding optimization and hardware-specific mechanics from programmers. We evaluate the system on controlled microbenchmarks against MP-SPDZ, focusing on online phase performance. Our CPU back end achieves up to $5.56\times$ speedup under intermediate and heavy algebraic workloads, shows strong scaling with thread count, and our GPU back end scales better as the input size increases. Overall, these results indicate that leveraging LLVM with protocol-aware scheduling is an effective architectural direction for extracting parallelism without sacrificing usability.
\end{abstract}

\section{Introduction}\label{sec:introduction}

Secure multiparty computation (MPC) enables a set of parties to jointly compute a function $f(x_0, x_1, \ldots)$ over their private inputs while revealing only the agreed output\cite{secure_computation_cite, pragmatic_mpc_cite}. A standard approach is secret sharing: each party splits its input into random-looking shares and distributes them so that no single party can reconstruct the input. Computation then proceeds on shares: additions and other linear operations are local, whereas non-linear operations, such as multiplication and comparison, invoke protocol-specific interactive subroutines. 

For a two-party sketch, let Alice hold $X$ and Bob hold $Y$, where $X$ and $Y$ denote their private inputs. They secret-share to obtain $[x_A], [x_B]$ and $[y_A], [y_B]$, giving Alice $[x_A], [y_A]$ and Bob $[x_B], [y_B]$. Because each share is individually indistinguishable from random, neither party learns the other’s input. After a sequence of local computations and brief interactions, each party obtains a share of $f(X,Y)$; combining these shares reveals only the output. 

We focus on optimizing actively secure arithmetic MPC for algebra-heavy workloads, targeting the SPDZ family \cite{spdz_cite}. MP-SPDZ \cite{mpspdz_cite} supports cross-protocol comparisons via a unified compiler and VM. However, its domain-specific language adds learning overhead, its blocking tape execution handles communication inefficiently, and it lacks a GPU back end for compute-intensive workloads. Piranha \cite{piranha_cite} offers a modular device layer and a C++ protocol API over secret-shared vectors/tensors, mapping batched operations to GPU kernels. This delivers strong speedups but expects vectorized inputs at the protocol boundary, shifting some batching responsibility to applications and complicating reuse across protocols.

MPC has real applications in privacy-preserving ML, cross-silo analytics, and private set operations such as Private Set Intersection (PSI) \cite{psi1_cite, psi2_cite}. Yet, large-scale deployments remain costly. Interactive steps, preprocessing, and integrity checks often shift the bottleneck to network costs, and privacy-preserving training and inference can still incur orders-of-magnitude overhead relative to cleartext implementations. \cite{privacypreservingml,securellm_cite}

To address these limitations, we developed an optimization pipeline based on LLVM~\cite{llvm_cite}, a mature compiler infrastructure with a static single assignment (SSA) intermediate representation (IR) \cite{llvm_cite}. 



We use LLVM IR to decouple front end semantics from protocol-specific execution, leveraging standard analyses and transformations to expose batching and parallelism in arithmetic MPC code. Specifically, we design and implement a proof-of-concept LLVM optimization pipeline for SPDZ that pairs compiler-guided batching with a non-blocking runtime that groups independent operations, dispatches batched messages for concurrent execution, overlaps communication with computation, and supports pluggable CPU and GPU back ends, mapping batches to GPU kernels when enabled.


Our evaluation shows that the LLVM-based pipeline delivers strong performance gains on the linear-layer microbenchmark. Under compute-heavy workloads, our CPU back end achieves up to $5.56\times$ speedup over MP-SPDZ. We also find that multithreading provides the largest benefits up to 8 threads for all systems, with our CPU back end exhibiting the strongest scaling. Finally, while our GPU back end is not competitive at the smallest inputs, it scales more favorably with input size, and its performance exceeds MP-SPDZ at larger input sizes. Overall, this yields a more portable, optimization-aware stack that 
leverages mature compiler infrastructure while keeping protocol mechanics isolated and maintainable.





In summary, our system achieves the following goals:
\begin{itemize}
\item \textbf{Leverage mature compiler infrastructure:} Provide a compilation stack based on the LLVM infrastructure to avoid error-prone re-implementation of analyses and transformations on the MPC programs.
\item \textbf{Isolate protocol-specific mechanics:} Keep protocol-dependent mechanics confined to well-defined layers to improve modularity and maintainability.
\item \textbf{Lower the programming barrier:} Support a mainstream front end language rather than a domain-specific language, reducing the learning curve for MPC developers.
\item \textbf{Shift optimization burden to the compiler/runtime:} Reduce the need for concurrency and hardware expertise by automatically applying batching, parallelization, and backend-specific tuning.
\item \textbf{Improve online performance:} Increase throughput by batching independent operations and overlapping computation with communication to mitigate communication overhead, with optional GPU offload for independent, compute-heavy workloads.
\end{itemize}
\section{Background}\label{sec:background}

This section briefly discusses some key concepts relevant to our ideas. It may be skipped by readers familiar with this material.

\subsection{SPDZ}

\subsubsection{Overview}
%
SPDZ is a general multi-party computation (MPC) protocol secure against an active adversary that can corrupt up to \(n-1\) of \(n\) parties \cite{spdz_cite}. It evaluates arithmetic circuits over a finite field. Execution is split into two phases: (i) an \textbf{offline} preprocessing phase that generates correlated randomness independent of the inputs (e.g., multiplication triples), and (ii) an \textbf{online} phase that consumes these preprocessed values, plus communication, to evaluate the circuit.

\subsubsection{Secret sharing and authentication}
Inputs and intermediate values are additively secret-shared so that, for each secret $x$,
\[
x \equiv \sum_{i=1}^n x_i \pmod p,
\]
and party \(P_i\) holds its share \(x_i\) \cite{secret_sharing_gen_cite,spdz_cite}. Any strict subset of the shares (fewer than \(n\)) reveals nothing about \(x\).
To achieve active security, SPDZ authenticates every share with an information-theoretic MAC under a \emph{global} secret key \(\alpha\) that is itself secret-shared. Party \(P_i\) holds a tuple \((x_i, m_i)\), where \(m_i\) is the MAC share for \(x_i\). When values are opened, the parties run a MAC check that verifies consistency under \(\alpha\) and detects cheating with statistical soundness.

\subsubsection{Secure operations}
We discuss two main secure operations: \textit{addition (local)} and \textit{multiplication (via Beaver triples)}.

\begin{itemize}
    \item \textbf{Addition (local).} Given shares of \(x\) and \(y\), each party outputs:
    \[
    z_i = x_i + y_i \pmod p,
    \]
    and MAC shares are added in the same fashion. No communication is required.

    \item \textbf{Multiplication (via Beaver triples).} Offline, the parties generate a tuple \((A,B,C)\) with \(C = A\cdot B\)\cite{beaver_triple_cite}, all secret-shared. At runtime, to multiply secrets \(x\) and \(y\):
    \begin{enumerate}[leftmargin=*]
      \item Each party computes \(d_i = x_i - A_i\) and \(e_i = y_i - B_i\); they open \(d=\sum_i d_i\) and \(e=\sum_i e_i\) (public).
      \item Each party locally reconstructs a share of \(z = x\cdot y\) using
      \[
      z = C + d\cdot B + e\cdot A + d\cdot e \pmod p,
      \]
      and updating the corresponding MAC shares. Only the masked values \(d\) and \(e\) are revealed; the triple is consumed and must not be reused.
    \end{enumerate}
\end{itemize}



\subsubsection{Security model}
SPDZ provides active (malicious) security in the dishonest-majority setting. If any party deviates, the MAC check fails and the protocol aborts, preventing leakage of secret information beyond what is explicitly opened.

\subsection{LLVM}\label{sec:llvm}

\subsubsection{LLVM Compiler Infrastructure}
LLVM is a collection of modular, reusable compiler and toolchain components 
(front end, middle end, and back end)
built around a common intermediate representation (IR)~\cite{llvm_project}. It is designed to support multiple source languages, allowing them to be lowered to the same IR, i.e., a machine-independent form of a program that encodes structure, control flow, and data dependencies. Compilers (e.g., \textit{Clang}~\cite{clang_project,rustc}) analyze IR and manipulate it as part of code optimization.

\subsubsection{Static Single Assignment (SSA) IR}
LLVM IR is in Static Single Assignment (SSA) form~\cite{cytron1991ssa,rodler2023ssa_phi}. SSA is a special form of traditional compiler Three-Address Code (TAC) IR~\cite{reiser1981threeaddr}, where each named value is defined exactly once, and each use refers to that unique definition~\cite{cytron1991ssa}. SSA makes the data dependencies (\textit{def–use chains}) explicit, simplifying many analyses and optimizations. 

\medskip
\noindent \textbf{Example.} Consider the following example to demonstrate the single assignment property of SSA. The source code is translated into SSA form on the right.

\begin{table}[h!]
\centering
\begin{tabular}{>{\raggedright\arraybackslash}m{0.3\linewidth} c >{\raggedright\arraybackslash}m{0.45\linewidth}}
\textbf{Source Code} & & \textbf{Static Single Assignment}\\
\begin{lstlisting}[language=C]
x = 0;
if (a > b) {
    x = a;
}
else {
    x = b;
}
return x;

\end{lstlisting}
&
&
\begin{lstlisting}[language=C]
x1 = 0
if (a1 > b1) {
    x2 = a1
}
else {
    x3 = b1
}
x4 = phi(x2, x3)
return x4
\end{lstlisting}
\end{tabular}
\end{table}

In the SSA representation, each variable is uniquely defined by a distinct versioned name (e.g., \texttt{x1}, \texttt{x2}, \texttt{x3}, and \texttt{x4}). The single-assignment property is preserved because every definition occurs exactly once. During execution, only one of the assignments to \texttt{x2} or \texttt{x3} is performed, depending on the outcome of the conditional expression \texttt{a1 > b1}. The $\phi$-function at the join point merges the two possible definitions and assigns the appropriate value to \texttt{x4}, reflecting the value of \texttt{x} after the conditional.

The general structure of SSA in LLVM IR can be expressed as shown below. Each variable (named with \%) is assigned exactly once, and control-flow merges are handled using $\phi$-nodes that select the appropriate value depending on the predecessor block.

\begin{lstlisting}[emptylines=1]
entry:
    %x1 = add i32 0, 0
    %cmp = icmp sgt i32 %a1, %b1
    br i1 %cmp, label %btrue, label %bfalse
btrue:
    br label %end
bfalse:
    br label %end
end:
    %x4 = phi i32 [ %a1, %btrue ], [ %b1, %bfalse ]
    ret i32 %x4
\end{lstlisting}

This form illustrates how LLVM represents control-flow and data-flow explicitly: conditional branches transfer control between labeled blocks, while the \texttt{phi} instruction merges values from different paths into a single SSA variable.



\subsubsection{Program structure and types}
LLVM organizes programs as modules containing functions; each function consists of basic blocks, i.e., straight-line sequences of typed SSA instructions ending in a terminator (e.g., branch or return). This organization yields a control-flow graph and a data-flow graph. The IR is strongly typed and provides pointer-based memory operations.



\subsubsection{Optimization passes}
LLVM’s optimizer includes a broad collection of IR-level analyses and transformations, such as constant propagation, dead-code elimination, global value numbering (GVN), loop optimizations, and vectorization, accessible through standard optimization pipelines (e.g., \texttt{-O2}, \texttt{-O3}) or user-defined pass sequences~\cite{llvm_passes_docs}. In this work, we leverage these protocol-independent optimizations to uncover parallel opportunities and to simplify redundant arithmetic before applying MPC-specific scheduling.

    
\section{Architecture}\label{sec:architecture}

\begin{figure}[t]
  \centering
  \includegraphics[width=\linewidth]{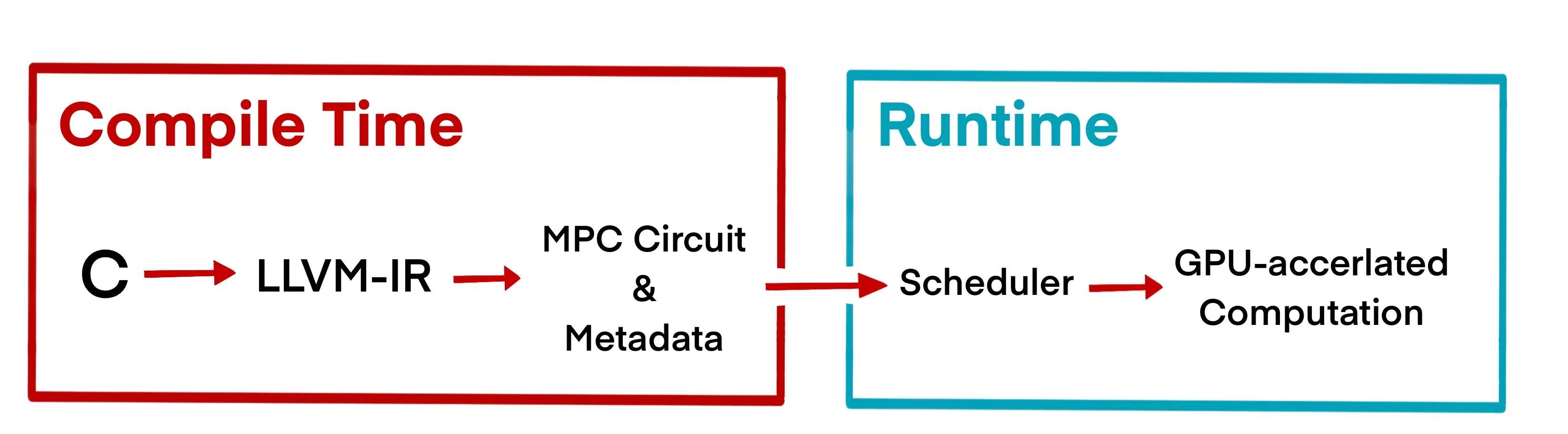}
  \caption{\textbf{Pipeline overview.} \emph{Compile time:} (1) annotated C is lowered to LLVM IR; (2) LLVM middle-end applies optimizations; (3) IR is converted into an MPC circuit via data-flow and control-flow analysis. \emph{Runtime:} (4) a non-blocking scheduler; (5) batchable regions are lowered to contiguous value/MAC buffers and executed on pluggable back ends (SIMD CPU or GPU).}
  \label{fig:pipeline}
\end{figure}

Figure~\ref{fig:pipeline} summarizes our end-to-end flow. At compile time, we (1) take an annotated C program and lower it to LLVM IR, preserving secrecy annotations and numeric domains; (2) apply LLVM middle-end optimizations (e.g., constant propagation, dead-code elimination, vectorization) to expose basic parallelism and reduce redundant arithmetic; and (3) analyze the IR’s data and control flow to recover MPC-relevant structure, such as \textit{def–use} chains, $\phi$-merges, and block/epoch boundaries, which we use to group instructions, linear or non-linear, into batchable regions, and create meta-datas to enable correct control-flow at runtime.
At runtime, the non-blocking scheduler orders these batchable regions while respecting data and control dependencies, overlapping heterogeneous work (local linear computations, interactive multiplies, opens/MAC checks) to reduce stalls. Each scheduled region is then lowered to contiguous value/MAC buffers and executed on a pluggable back end (SIMD CPU or GPU) through a narrow interface.

\subsection{Translation from high level to LLVM IR}

\begin{figure*}[t]
  \centering
\begin{lstlisting}[language=C,basicstyle=\ttfamily\small,frame=single]
#include <stdint.h>

uint32_t* linear_layer(const uint32_t *x __attribute__((annotate("private"))),
                       const uint32_t *W __attribute__((annotate("private"))),
                       const uint32_t *b __attribute__((annotate("public")))) {
    uint32_t *out = mark_linear_layer(x, W, b, DIN, DOUT);
    return out;
}
\end{lstlisting}
  \caption{\textbf{Example C front end with privacy annotations}
  Parameters are annotated as \texttt{private} or \texttt{public}, which propagates to LLVM IR.
  The reserved call \texttt{mark\_linear\_layer} marks an ML linear layer ($f = W x + b$) for
  protocol-aware handling. Programmers may also implement the same computation with ordinary loops;
  annotations alone suffice for correctness.}
  \label{fig:c-interface}
\end{figure*}

\begin{figure*}[t]
  \centering
\begin{lstlisting}[basicstyle=\ttfamily\small,frame=single]
 Vectorized LLVM IR (SSA). Example loads and adds with width = 128
  %18 = getelementptr inbounds i32, ptr %0, i64 %17
  %19 = load <128 x i32>, ptr %18, align 4, !tbaa !6
  %20 = getelementptr inbounds i32, ptr %18, i64 128
  %21 = load <128 x i32>, ptr %20, align 4, !tbaa !6
  %22 = getelementptr inbounds i32, ptr %18, i64 256
  %23 = load <128 x i32>, ptr %22, align 4, !tbaa !6
  %24 = getelementptr inbounds i32, ptr %18, i64 384
  %25 = load <128 x i32>, ptr %24, align 4, !tbaa !6
  %26 = add <128 x i32> %19, %13
  %27 = add <128 x i32> %21, %14
  %28 = add <128 x i32> %23, %15
  %29 = add <128 x i32> %25, %16
\end{lstlisting}
  \caption{\textbf{A fragment of vectorized LLVM IR produced after front end compilation and middle-end passes.}
  Loads use 128-bit wide vectors (\texttt{<128 x i32>}); elementwise adds are fused into
  vector operations (\texttt{\%26--\%29}). Such width-driven IR is a first source of batching:
  it compacts many homogeneous element-wise operations into a small number of back end primitives.}
  \label{fig:llvm-ir}
\end{figure*}

Our pipeline starts at the user-facing front end. The programmer writes an MPC routine in any language that can be compiled to LLVM IR (e.g., C/C++, Rust). The MPC entry point is a function that takes user inputs as parameters and returns the value the parties agree to reveal. Our implementation supports a practical subset of C: arithmetic and logical operators, array indexing, for/while loops, if statements, etc. The only extra requirement is a lightweight privacy annotation on parameters (e.g., private vs. public). These annotations propagate into the IR, and later IR analyses ensure that any computation data-dependent on private inputs is routed through the correct secure sub-protocols.

Figure~\ref{fig:c-interface} shows a minimal C example:
x and W are annotated private, while b is public.
The call to mark-linear-layer(...) is a reserved “black-box” marker that identifies an ML linear layer (f = W x + b) to the compiler.

We provide such black-box markers for certain compound operations (e.g., linear layers). Treating them as single, protocol-aware operations lets us apply MPC-specific optimizations (e.g., Beaver-triple partitioning) more aggressively. Importantly, programmers are not required to use these markers; the same functionality can be written in plain C using loops and array operations without sacrificing parallelism. Doing so may leave some protocol-specific optimizations on the table, but it significantly lowers the entry cost for users who prefer standard C without learning new domain-specific language constructs or optimization pragmas.

Because the front end is a mainstream language, users benefit from familiar tooling: editors, debuggers, sanitizers, and profilers all work as usual. The programmer focuses on correct semantics and privacy annotations; LLVM's \texttt{clang} compiler takes care of producing IR with standard optimizations enabled.

Once the C program is written with privacy annotations for each input, we invoke \texttt{clang} to obtain the textual LLVM IR used by the rest of our pipeline. For example, for our linear-layer test, we use:
\begin{lstlisting}[language=bash,basicstyle=\ttfamily\small]
clang -S -O2 \
  -DDIN={N} -DDOUT={M}} \
  -mllvm -force-vector-width={vec_len}} \
  -mllvm -force-vector-interleave={interleave_len} \
  -emit-llvm input.c \
  -o output.ll
\end{lstlisting}
This produces an \texttt{.ll} file in LLVM's SSA form. Preprocessor defines (\texttt{DIN}, \texttt{DOUT}) as input dimensions; \texttt{-O2} enables LLVM's standard middle-end pipeline; \texttt{-force-vector-width} specifies the target SIMD width (number of lanes) for vectorized loops, while \texttt{-force-vector-interleave} sets the interleave factor. As an exmaple, these flags help produce the vectorized IR shown in Figure~\ref{fig:llvm-ir}.




We rely on a small set of LLVM optimizations that are particularly important for MPC. Vectorization groups multiple scalar loop iterations into a single wide instruction (e.g., \texttt{<4 x i32>} loads and adds), which directly matches our batched arithmetic primitives and reduces per-element overhead. Constant propagation and dead-code elimination simplify expressions and remove computations whose results are never used, shrinking the circuit. Loop-invariant code motion (LICM) hoists computations that do not depend on the loop index out of loops, so they are not redundantly recomputed inside the MPC circuit. Global value numbering (GVN) identifies equivalent expressions and reuses their results instead of recomputing them, which reduces both arithmetic cost and communication.

These optimizations are protocol-agnostic but directly valuable in the MPC setting because they reduce the number of arithmetic operations and memory accesses that must be performed on secret shares. Reusing LLVM’s mature optimizer shifts most tuning effort from programmers to the compiler: standard \texttt{-O2}/\texttt{-O3} pipelines pick profitable unroll, vector-width, and interleave factors based on loop structure and target architecture\cite{clang-cmdline-ref}. Our later MPC-specific passes can therefore focus on secrecy-aware scheduling and batching, assuming that the input IR is already simplified, vectorized where profitable, and expressed in SSA form with clear def--use structure, which we will discuss in more detail in the next subsection.


\subsection{Dataflow and Control flow Analysis}
After lowering to LLVM IR, we build two tightly coupled views of the program: a custom data-flow graph (DFG) that encodes data dependency between operations, and a control-flow graph (CFG) that encodes the legal execution order. A data-flow graph is a directed graph whose nodes are computation steps and whose edges indicate that the value produced by one step is consumed by another. A classical def--use chain for a variable connects a \emph{definition} (where the variable is assigned) to all \emph{uses} (where it is read) that are reachable without an intervening redefinition. Because LLVM IR is in SSA form, each SSA value has exactly one definition, so building def--use chains reduces to simply connecting each SSA definition to its syntactic uses. This pairing lets the scheduler use the DFG to pick kernels and batch work, while using the CFG to respect program order, handle joins, and advance execution safely.

\subsubsection{Data flow — extracting an MPC-friendly graph from LLVM IR}\label{lbl:dfg-analysis}

\begin{figure}[t]
  \centering
  \begin{tikzpicture}[
      >=stealth,
      node distance=0.8cm and 0.6cm,
      var/.style={circle,draw,minimum size=0.7cm,inner sep=1pt},
      lbl/.style={font=\ttfamily\small}
    ]

    \node[lbl] (inst) at (0,1.6)
      {\texttt{\%3 = add i32 \%1, \%2}};

    \node[var] (add) at (0,0) {Add};

    \node[lbl] (v3) [above=2pt of add] {\texttt{\%3}};

    \node[var] (v1) [below left=of add] {\texttt{\%1}};
    \node[var] (v2) [below right=of add] {\texttt{\%2}};

    \draw[->] (v1) -- (add);
    \draw[->] (v2) -- (add);
  \end{tikzpicture}
  \caption{\textbf{Example DFG for a basic block with an addition.}
  LLVM instructions of the form \texttt{\%3 = op(...)} become DFG nodes whose
  incoming edges are the operand values and whose outgoing edge is the defined SSA value.}
  \label{fig:dfg-add-example}
\end{figure}
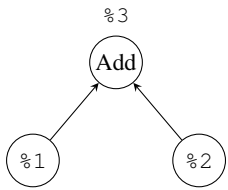

We start from LLVM’s SSA IR and build a sparse \emph{data-flow graph (DFG)} in which every SSA definition becomes a node and edges connect each definition to its uses. Concretely, we scan each instruction of the form \texttt{\%id = op(arg\_1, ..., arg\_n)}, create a node for \texttt{\%id}, and wire its operand IDs as incoming edges. For example, an instruction
\begin{lstlisting}[basicstyle=\ttfamily\small]
%3 = add i32 %1, %2
\end{lstlisting}
becomes a node \texttt{Adder(\%1, \%2) -> \%3} in our graph, with edges from the nodes for \texttt{\%1} and \texttt{\%2} into the node for \texttt{\%3}. Our DFG differs slightly from typical compiler DFGs: in our implementation, operator nodes are parents and operand values are children, but the semantic meaning of the dependencies remains the same. Figure~\ref{fig:dfg-add-example} illustrates our translation on a small basic block.

Instructions that do not define a new SSA value (e.g., \texttt{store} or debug intrinsics) either have no direct counterpart in the MPC circuit or are used only to determine which values become outputs. We therefore treat them as control or metadata rather than first-class arithmetic nodes: their effects are reflected indirectly through the \texttt{Load} and \texttt{ret} nodes described below, rather than as separate gates in the MPC graph.

Compile-time constants are also represented as nodes with stable IDs (interned via a small constant pool). The SSA value returned by \texttt{ret \%v} is recorded as the graph \emph{root}—its availability marks completion of the computation. We also encode some control-flow information into the DFG, which we will explain further in Section~\ref{lbl:cfg-analysis}.

\paragraph{Normalizing parameter loads}
Inputs that arrive through pointer parameters are converted into explicit \texttt{Load(base, start, count)} nodes in the DFG. Each such node has three children:
\begin{itemize}
  \item \texttt{base} is the ID of the input parameter (found by following trivial aliases and the final \texttt{getelementptr} (GEP) chain back to the parameter),
  \item \texttt{start} is the accumulated index expression from the last GEP in that chain,
  \item \texttt{count} equals the IR type’s vector width (e.g., $N$ for \texttt{<N x i32>}, or $1$ for scalars).
\end{itemize}
Conceptually, \texttt{Load(base, start, count)} is a single node that represents ``read \texttt{count} consecutive elements starting at offset \texttt{start} from the buffer identified by \texttt{base}''. For instance, a sequence
\begin{lstlisting}[basicstyle=\ttfamily\small]
%ptr  = getelementptr inbounds i32, ptr %x, i64 %idx
%val  = load <4 x i32>, ptr %ptr
\end{lstlisting}
becomes a node \texttt{Load(x, idx, 4)} whose outgoing edge is the vector value \texttt{\%val}. These three children encode public metadata about where and how much to read; the loaded values themselves may, of course, be secret after sharing. This normalization gives the runtime precise, sliceable views of inputs/weights/biases and lets it batch I/O deterministically.

\paragraph{Vectorization is first-class}
The graph preserves lane width as part of the node kind. Scalar integer ops become \texttt{Adder}, \texttt{Multiplier}, \texttt{Subtract}; vector ops become \texttt{AddBatch}, \texttt{MultBatch}, \texttt{SubBatch} with lane counts implied by type. Horizontal vector intrinsics such as \texttt{llvm.vector.reduce.add} lower to \texttt{ReduceAdd(vec)} (and analogously for multiply). Because widths and reductions are explicit, the runtime doesn’t need to re-infer shapes and can choose element-wise kernels vs.\ reduction trees directly from the node kinds.

\paragraph{Lowering IR idioms to an MPC gate set}
We restrict the DFG to arithmetic (field) operations that map cleanly to secret sharing:
\begin{itemize}
  \item Basic \texttt{add/mul/sub} map directly (scalar or batch).
  \item Comparisons(\texttt{<}, \texttt{>}, etc.) are canonicalized. Equality/inequality are expressed arithmetically as 0 or 1, eliminating the need for boolean gates.
  \item \texttt{select(c,t,f)} is normalized to arithmetic as $f + c\cdot(t-f)$, preserving semantics while staying in the arithmetic circuit domain.
  \item \texttt{shl} by an immediate $k$ lowers to multiplication by $2^k$.
  \item Zext are treated as pass-throughs (input bits and range are fixed).
  \item Bitwise idioms(AND, OR, XOR) are converted to its arithmetic equivalent.
  \item Domain-specific fast path: calls to the front end hook \newline \texttt{mark\_linear\_layer(x, W, b, DIN, DOUT)} are collapsed into a single \texttt{LinearLayer(x, W, b)} node carrying explicit sizes. This makes the operator share visible to the runtime and preserves the high-level structure for triple pre-fetching and GPU kernels.
\end{itemize}

\paragraph{Seeding and propagating privacy}
We read \texttt{@llvm.var.annotation} to tag function parameters as \emph{private} or \emph{public}, attach these tags to the corresponding \texttt{Input} nodes, and propagate privacy sparsely along def$\to$use edges in later passes. Any value that depends on a private input becomes private. This lightweight tainting enables secrecy-aware kernel selection and MAC-check placement at runtime without re-parsing IR.

\subsubsection{Compile-time Control-Flow Analysis}\label{lbl:cfg-analysis}

At compile time, we analyze LLVM basic blocks and branch instructions and produce a canonical control-flow graph (CFG) plus minimal metadata that the runtime will later consult to enforce legal execution order. We also encode control-relevant information in the MPC data-flow graph (DFG). (How the scheduler uses these artifacts and why control-flow handling is needed appears in Section~3.3.2.)

LLVM IR already provides a CFG structure: functions are partitioned into basic blocks, and a terminator instruction (conditional or unconditional branches) are always present as the last instruction of each block to identify successors. We reuse this structure directly and transform it into an MPC-friendly format rather than reconstructing it from scratch.

We first construct a \texttt{BlockLabel} node for every LLVM basic block. A \texttt{BlockLabel} exists purely for control purposes: it denotes ``entry to block $B$'' and serves as the anchor for all nodes that correspond to IR instructions in that block. In the DFG, each instruction node carries a pointer to its owning \texttt{BlockLabel}, so the scheduler can relate data dependencies back to block-level control. PHI nodes themselves remain ordinary data nodes in the DFG (fed by predecessor labels recorded in their operands). Branch instructions (conditional or unconditional) are also present as control-flow purpose nodes in both CFG and DFG.

Our custom CFG has two layers:
\begin{itemize}[leftmargin=1.2em]
  \item \textbf{Intra-block linear order.} Inside each block, we canonicalize a single linear chain that mirrors LLVM's instruction order after our IR lowerings. The corresponding \texttt{BlockLabel} points to the leader of this instruction chain in the DFG, making it the natural entry point for the block.
  \item \textbf{Inter-block control transfers.} Inter-block edges come only from \texttt{Branch} nodes. Unconditional branches have a single successor (the target \texttt{BlockLabel}); conditional branches have two successors (true/false \texttt{BlockLabel}s). In the finalized CFG, \texttt{Branch} successors are \texttt{BlockLabel}s only.
\end{itemize}

We also emit lightweight loop metadata so the runtime can advance through loop regions safely---even with nested loops---without reanalyzing IR. We treat a \texttt{BlockLabel} as a loop header if it has a later block that branches back to it (a back edge). For each header we compute: (i) the set of \emph{loop members} (blocks reachable from the header that can also reach the header), and (ii) the set of \emph{loop exits} (successors that leave this member set). These two lists give the runtime scheduler explicit region boundaries and exit points without reanalyzing the IR, and they are all attached to the corresponding \texttt{BlockLabel} nodes as metadata. We will explain how this metadata is used to enforce correct control flow in the next section.


\subsection{Runtime Scheduling}

\subsubsection{Overview}
At runtime, the scheduler treats the program as the conjunction of a data-flow graph (DFG) and a control-flow graph (CFG). A node becomes eligible to execute only when two preconditions are simultaneously satisfied: (i) all data dependencies have produced values, and (ii) control has reached the node’s basic block along some taken path. This “data-and-control readiness” rule prevents speculative execution across untaken branches and ensures correctness in the presence of public control decisions and MPC secrecy constraints.

\subsubsection{Readiness of operations}
Parameters and constants are treated as immediately available leaves; readiness then propagates along DFG edges as producers complete. Within loops, the scheduler distinguishes \emph{stable} inputs (originating outside the current loop iteration) from \emph{iterative} inputs. Stable inputs are considered satisfied for the entire iteration, so only iterative producers gate per-iteration progress. On the control side, a basic block becomes executable only after control transfers to its label; until then, even data-ready operations in the block are withheld. Upon block entry, any $\phi$-nodes are resolved by predecessor: the taken branch records a predecessor label, and each $\phi$ selects the value associated with that label, preserving SSA semantics without introducing secret control.

\subsubsection{Non-blocking scheduler loop}
The scheduler never pauses for global barriers. Instead, it repeatedly asks for the next ready node and issues work whenever either data or control becomes available, overlapping independent compute with long-latency cryptographic steps. Algorithm~\ref{alg:scheduler} summarizes this core loop. The scheduler maintains two ready queues: one for cheaper add-like operations and one for more expensive mul-like operations. It greedily prefers the $\textit{ready\_heavy\_ops}$ queue when non-empty to expose longer-latency work early, but otherwise falls back to $\textit{ready\_light\_ops}$. This decision puts higher-latency, communication-heavy operations in flight earlier, so their cost can be overlapped with cheaper add-like work that is issued later. Branches are treated specially: rather than being executed by the back end, they are processed in-place via \textsc{TryBranchOnce}, which may enter new blocks, seed $\phi$-nodes, and inject additional ready nodes into the queues. If all remaining items are stalled branches, the scheduler returns $\bot$ and the runtime can wait for in-flight work to complete.

\begin{algorithm}[h!]
\small
\caption{Non-blocking MPC scheduler loop}
\label{alg:scheduler}
\DontPrintSemicolon
\SetKwProg{Fn}{Function}{}{end}

\Fn{NextReadyNode()}{
  \While{true}{
    \If{$\textit{ready\_heavy\_ops}$ not empty}{
      pop $id$ from $\textit{ready\_heavy\_ops}$\;
      \Return{$(id, \textsc{NodeType}(id), \textsc{Children}(id))$}\;
    }
    \If{$\textit{ready\_light\_ops}$ empty}{
      \Return{$\bot$} \tcp*[r]{no compute work available}
    }
    pop $id$ from $\textit{ready\_light\_ops}$\;
    \eIf{$\textsc{NodeType}(id)$ is \textsc{Branch}}{
      \If{\textsc{TryBranchOnce}$(id)$}{
        \tcp{branch taken: control moved to successor block(s), PHIs seeded,}
        \tcp{and any newly ready nodes enqueued; loop to pick next compute node}
      }
      \Else{
        requeue $id$ into $\textit{ready\_light\_ops}$ \tcp*[r]{branch not yet allowed by control flow}
      }
    }{
      \Return{$(id, \textsc{NodeType}(id), \textsc{Children}(id))$}
    }
  }
}
\end{algorithm}

\subsubsection{Parallelism vs.\ control flow}
Block structure does not limit available parallelism. For acyclic (non-loop) regions, multiple basic blocks can be reachable and active at the same time; once a block is entered, any of its data-ready operations may run concurrently with ready operations in other entered blocks. Thus, block boundaries serve only to prevent premature execution, not to serialize work. Loops introduce the only necessary synchronization: an iteration may repeat (take the back-edge) only after the current iteration’s body has completed. This avoids cross-iteration races while still permitting maximal overlap within an iteration and across unrelated code outside the loop. Overall, the non-blocking scheduler in Algorithm~\ref{alg:scheduler} keeps the pipeline busy and masks latency from cryptographic communication and Beaver-triple usage, while respecting control-flow constraints.

\subsubsection{Loop handling and back-edge gating}
If LLVM fully unrolls a loop, it appears as straight-line code and executes like any other acyclic region. Otherwise—when a loop is too large to unroll or its trip count is runtime-known—the IR contains a loop header and one or more blocks that eventually branch back to that header. Such runtime loops are governed by per-iteration epochs:
\begin{itemize}
\item \textit{Start of iteration.} Entering the header advances the epoch and marks the beginning of a new iteration.
\item \textit{Progress tracking.} During an epoch, the scheduler tracks which operations inside the loop’s region have completed.
\item \textit{Back-edge gating.} A back-edge may be taken only when a CFG walk from the header to that back-edge is covered by operations already completed in the current epoch. The walk is confined to the loop’s region, so code outside the loop does not constrain iteration progress.
\item \textit{Nested loops.} When the walk encounters an inner loop, the scheduler treats it compositionally: it reasons at the granularity of the inner loop’s exits for the current outer epoch, rather than expanding all inner iterations.
\end{itemize}
Algorithm~\ref{alg:loop-gating} gives the corresponding implementation of \textsc{TryBranchOnce} and \textsc{IsLoopEpochComplete}. Before a back-edge is taken, \textsc{IsLoopEpochComplete} checks that one CFG path from the loop header to the candidate branch has all arithmetic nodes stamped as completed in the current epoch, treating nested loops via their exits. If the branch exits the loop instead of closing the back-edge, the loop is marked finished and its epoch is set to a sentinel value.

\begin{algorithm}[h!]
\small
\caption{Loop gating for back-edges}
\label{alg:loop-gating}
\DontPrintSemicolon
\SetKwProg{Fn}{Function}{}{end}

\Fn{TryBranchOnce($id$)}{
  $(children) \gets$ operands of $id$; \quad
  $curBlk \gets$ block containing $id$ (if any)\;
  \eIf{branch is unconditional}{
    $dst \gets$ unique successor label from $children$\;
  }{
    read public scalar condition $c$ and choose $dst$ (true / false)\;
  }
  $H \gets$ \textsc{LoopHeaderForBranch}$(id)$\;
  \If{$H \neq \bot$}{
    $e \gets loop\_epoch[H]$\;
    \If{$e > 0$ \text{ and } \textsc{IsLoopEpochComplete}$(H, id, e)$ is \textbf{false}}{
      \Return{\textbf{false}} \tcp*[r]{loop body not done for epoch $e$}
    }
    \If{$dst \neq H$}{
      \textsc{MarkLoopFinished}$(H)$\;
    }
  }
  \If{$curBlk$ exists}{
    \textsc{WritePredecessorLabelValue}$(dst, curBlk)$\;
  }
  \textsc{EnterBlock}$(dst)$; \quad $node\_state[id] \gets base\_unlock[id]$\;
  \Return{\textbf{true}}\;
}

\Fn{IsLoopEpochComplete($H, branch, e$)}{
  $members \gets$ blocks in loop region of $H$; \quad
  worklist $\gets$ CFG successors of $H$; \quad $seen \gets \emptyset$\;
  \While{worklist not empty}{
    pop node $u$ from worklist\;
    \If{$u \in seen$}{\Continue}
    insert $u$ into $seen$; \quad
    $blk \gets$ block containing $u$ (if any)\;
    \If{$blk \notin members$ \text{ and } $blk \neq H$}{\Continue}
    \If{$u = branch$}{\Return{\textbf{true}}}
    \If{$u$ is header of a nested inner loop}{
      \ForEach{exit label of this inner loop}{
        push successors after that exit into worklist\;
      }
      \Continue
    }
    \If{$u$ is an arithmetic node in $H$'s loop}{
      $done \gets last\_done\_epoch[(H, u)]$ (default $0$)\;
      \If{$done < e$}{\Return{\textbf{false}}}
    }
    \If{$u$ is a conditional branch in the loop and $u \neq branch$}{
      read public condition and push successors after the taken label\;
    }
    \Else{
      push all CFG successors of $u$ into worklist\;
    }
  }
  \Return{\textbf{false}}\;
}
\end{algorithm}

\subsubsection{Control flow must be public}
Conditional branches are permitted only on public conditions; if a condition is secret, attempting to steer control flow is rejected. For public conditions, the scheduler resolves the taken successor, records the predecessor label for the destination block, and performs block-entry actions (including seeding $\phi$-nodes and applying deferred dependencies). Unconditional branches are handled analogously, without the guard. In both cases, loop gating from Algorithm~\ref{alg:loop-gating} constrains branches that would close a loop iteration.

\subsubsection{Structured Data Movement}
Indexing and slicing operations execute as soon as their scalar parameters are available and produce views onto existing data rather than actual copies. This keeps scheduling orthogonal to layout and reduces needless data motion across CPU/GPU back ends, which we will discuss further in Section~3.4.

\subsection{Data partitioning and ML}\label{data-partitioning}
We implemented LinearLayer, the classic $y = W x + b$ operator, as a first ML workload because it is both widely used and computationally heavy. Dense matrix--vector (and matrix--matrix) products dominate many inference/training pipelines, map cleanly to batched, vectorized kernels, and exercise exactly the resources our runtime and supported GPU back end are designed to exploit: wide parallelism, high arithmetic intensity, and predictable memory access.

\subsubsection{Overview}
We treat the linear layer \(y = W x + b\) as a first-class operator with ML-specific partitioning, while exposing only a uniform black-box interface to the scheduler. Here \(\mathrm{DIN}\) is the input length (\(\lvert x\rvert\)) and \(\mathrm{DOUT}\) is the output length (\(\lvert b\rvert\)), with \(\lvert W\rvert\) = \(\mathrm{DIN} \times \mathrm{DOUT}\). This design keeps memory usage predictable, maximizes kernel throughput, and reduces cross-party rounds without leaking shape details into scheduling logic. The black-box call avoids expanding the layer into a deep circuit and enables MPC-specific optimizations. At the same time, a plain C implementation that does not call this operator is still optimized by the compiler: standard LLVM passes and our data-flow analysis vectorized and parallelize the code, so most benefits remain even without the specialized entry point.

\subsubsection{Data slicing}
\noindent When input size is large, we tile \(W\) by contiguous row blocks and pair each block with the matching slice of \(b\). The full input vector \(x\) is reused across all tiles. Each tile is represented as a single \texttt{ShareBatch} view (values + MACs). Tile size is tunable: larger tiles increase arithmetic intensity and reduce per-tile overhead; smaller tiles lower latency and improve pipeline depth. The scheduler issues tiles in a non-blocking way and consumes whichever results arrive first, reducing idle time and smoothing load across CPU/GPU workers.

\subsubsection{MPC-specific optimization}
\noindent For private \(x\) and \(W\), we use matrix triples \((A,B,C)\) with \(C=A\cdot B\) at tile granularity instead of many scalar triples. For a tile with operands \((X,Y)\), parties compute and open \(D=X-A\) and \(E=Y-B\), and then follows the standard batched MAC check path to verify integrity. In parallel, the product share for the tile is reconstructed using the usual Beaver identity:
\[
Z \;=\; C \;+\; D\cdot B \;+\; A\cdot E \;+\; D\cdot E.
\]
We allocate one triple per tile. This aligns exactly with the row-tiling layout, amortizes preprocessing over an entire tile instead of per-element multiplies. Triple shapes match the tile size, so a single preprocessing pool can flexibly serve different tile sizes without regeneration.

\section{Evaluation}\label{sec:evaluation}

We evaluate our approach with a prototype that supports both CPU and GPU back ends, and compare against MP-SPDZ using a controlled microbenchmark based on a single linear layer. We emphasize online-phase performance and study how runtime varies with thread count, number of parties, slice size, and input size. All experiments run on a single-GPU machine, so GPU results may reflect contention; we leave multi-GPU evaluation to future work. For more information, see Section \ref{gpu_performance}.

\subsection{Experimental Setup}
All experiments were conducted on the cluster's GPU partition with access to one NVIDIA RTX A6000, 64 CPU cores, and 256 GB of memory.

We use the cluster module system to keep compiler and library versions consistent across runs. For MP-SPDZ, we load a GCCcore toolchain and standard dependencies (CMake, GMP, Boost, Python, and CUDA) before building and running the benchmarks.

The following parameters are fixed across all systems/frameworks in this evaluation:
\begin{itemize}
    \item Input bits: 32.
    \item Prime field: $p = 4294967291$ ($2^{32} - 5$).
    \item GPU threads per block: 1024.
    \item When the GPU back end is enabled, operations with vector size below \texttt{MIN\_KERNEL\_SIZE} fall back to CPU execution to avoid kernel launch overhead. We set \texttt{MIN\_KERNEL\_SIZE} $= \texttt{slice\_size}/4$.
    \item Since we are mainly targeting the online phase performance, we evaluate both our system and MP-SPDZ using fake preprocessing.
\end{itemize}

\subsection{Benchmark Workload}

All tests target a single ML linear layer as described in Section~\ref{data-partitioning} . The program takes a private input vector, a private matrix, and a private bias vector; all inputs are secret-shared before computation.

Our experiments aim to answer the following research questions:
\begin{enumerate}
    \item How does performance scale with the number of threads per party under a fixed workload?
    \item How does performance scale with the number of parties under a fixed workload?
    \item How does slice size affect performance, and how does it interact with thread count?
    \item How does performance scale with input size?
    \item What is the breakdown across front end, setup, and online phases?
\end{enumerate}




\subsection{RQ1: Thread Scaling}\label{subsec:thread-scaling}

Multithreading is a core component of our system. It enables parallel execution of independent operations and supports non-blocking scheduling of the computation. We measure thread scalability by comparing our SPDZ implementation on the CPU and GPU back ends against MP-SPDZ while varying the thread count under a fixed workload.

\begin{figure}[t]
    \centering
    \small
    \begin{tabular}{ll}
        \hline
        \textbf{Setting} & \textbf{Value} \\
        \hline
        Input/Output dimension / size & 8k/64m \\
        Parties & 2 \\
        Slice size (ours) & 262k \\
        \# Computation threads per party (X-axis) & \{1, 2, 4, 8, 16, 32, 64\} \\
        \hline
    \end{tabular}

    \vspace{0.6em}

    \includegraphics[width=0.95\linewidth]{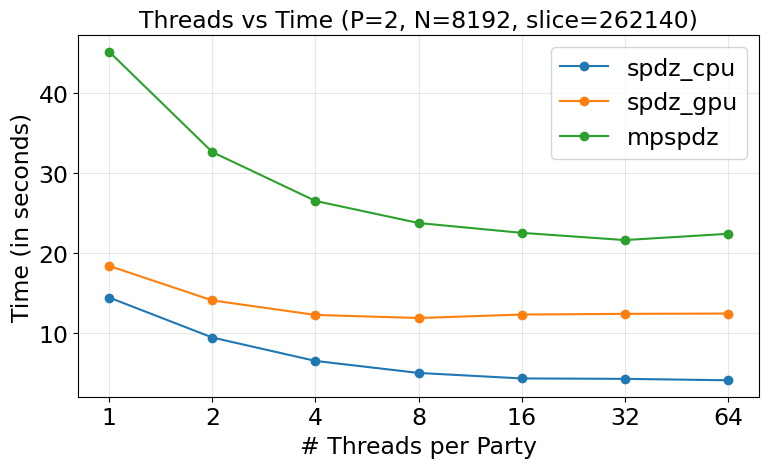}
    \caption{Thread scaling comparison between our SPDZ implementation (CPU/GPU) and MP-SPDZ. }
    \label{fig:thread-scaling}
\end{figure}

We fix the workload to a single linear layer with Input/Output dimension/size = 8192/67108864, and we fix the number of parties to 2. For our implementation, we additionally fix the slice size to 262140 when thread count is the only variable. MP-SPDZ does not have a directly configurable slice size. We use the same thread set for all systems. The resulting performance curves are shown in Figure~\ref{fig:thread-scaling}.

As shown in Figure~\ref{fig:thread-scaling}, all three curves show the largest reductions in runtime when increasing threads from 1 to 8. MP-SPDZ decreases from 45.21s at 1 thread to 23.74s at 8 threads. Our CPU back end decreases from 14.40s to 4.95s over the same range, and our GPU back end decreases from 18.36s to 11.83s. Beyond 8 threads, changes are smaller for all systems: between 8 and 64 threads, MP-SPDZ varies within 21.61--23.74s, our GPU back end within 11.83--12.39s, and our CPU back end within 4.02--4.95s.

Across the full thread set, our CPU back end reports the lowest runtime at each thread count, followed by our GPU back end, with MP-SPDZ highest for this configuration. At 1 thread, the measured times are 14.40s (CPU), 18.36s (GPU), and 45.21s (MP-SPDZ). At 8 threads, they are 4.95s, 11.83s, and 23.74s, and at 64 threads, 4.02s, 12.39s, and 22.39s, respectively.

\subsubsection{Multithreading gains and cross-system speedups}
Multithreading benefits all three systems, but the magnitude differs substantially. Relative to the single-thread baseline, MP-SPDZ achieves up to a $2.09\times$ speedup (45.21s $\rightarrow$ 21.61s), corresponding to a 52.2\% reduction in runtime at 32 threads. Our CPU back end scales more strongly, reaching a $3.58\times$ speedup (14.40s $\rightarrow$ 4.02s), or a 72.0\% reduction, at 64 threads. Our GPU back end shows a smaller peak gain of $1.55\times$ (18.36s $\rightarrow$ 11.83s), a 35.6\% reduction at 8 threads, after which improvements flatten.

Compared against MP-SPDZ at the same thread counts (2 parties, Input/Output dimension/size = 8192/67108864), our CPU back end provides consistent and growing advantages: $3.14\times$ faster at 1 thread, $4.80\times$ faster at 8 threads, and $5.56\times$ faster at 64 threads (68.2\%, 79.2\%, and 82.0\% lower runtime, respectively). Our GPU back end is $2.46\times$ faster than MP-SPDZ at 1 thread and remains about $1.75$--$2.01\times$ faster for 8--64 threads (roughly 43--50\% lower runtime).

\subsection{RQ2: Party Scaling}\label{subsec:party-scaling}

We evaluate how performance changes as the number of parties increases under a fixed workload. As the party count grows, the communication pattern becomes denser; in many multi-party protocols, the number of pairwise communication channels scales on the order of $P^2$, which can increase total communication volume and synchronization overhead. 

\begin{figure*}[t]
    \centering
    \small

    \begin{tabular}{ll}
        \hline
        \textbf{Setting} & \textbf{Value} \\
        \hline
        Input/Output dimension / size & 8k/64m \\
        Slice size (ours) & 262k \\
        Parties (X-axis) & \{2, 3, 4, 5, 6\} \\
        \# Computation threads per party & $T \in \{1, 8, 32\}$ \\
        \hline
    \end{tabular}

    \vspace{0.6em}

    \begin{subfigure}[t]{0.32\textwidth}
        \centering
        \includegraphics[width=\linewidth]{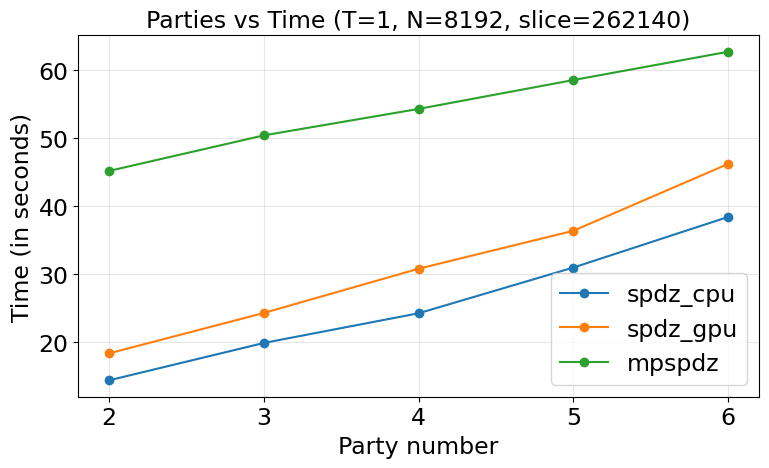}
        \caption{$T=1$}
        \label{fig:party-scaling-t1}
    \end{subfigure}
    \hfill
    \begin{subfigure}[t]{0.32\textwidth}
        \centering
        \includegraphics[width=\linewidth]{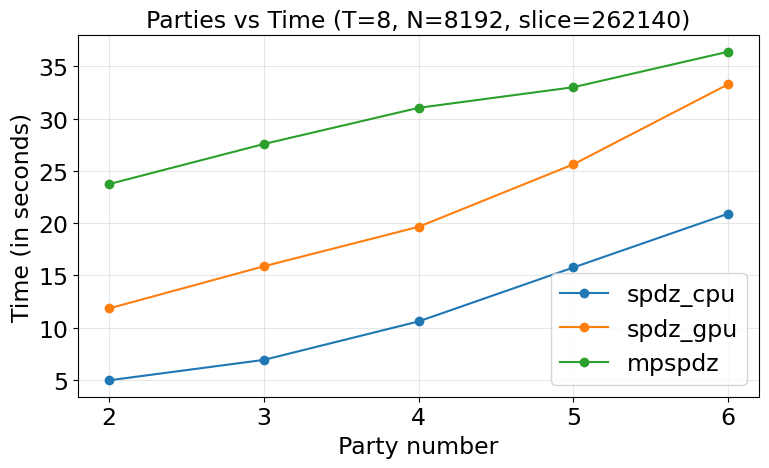}
        \caption{$T=8$}
        \label{fig:party-scaling-t8}
    \end{subfigure}
    \hfill
    \begin{subfigure}[t]{0.32\textwidth}
        \centering
        \includegraphics[width=\linewidth]{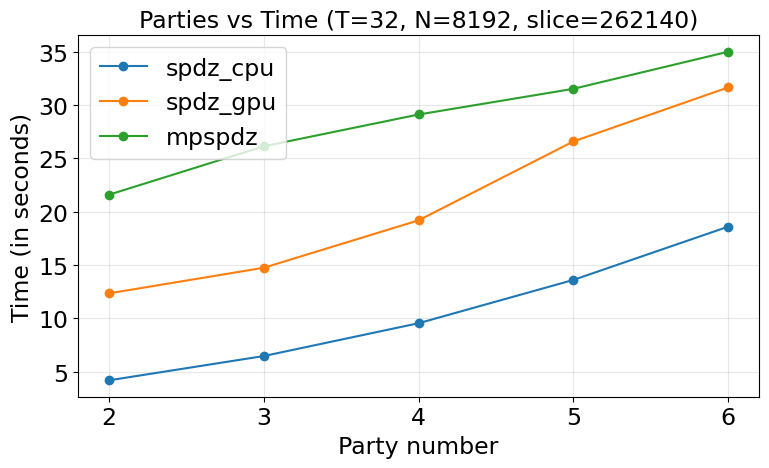}
        \caption{$T=32$}
        \label{fig:party-scaling-t32}
    \end{subfigure}

    \caption{Party scaling comparison between our SPDZ implementation (CPU/GPU) and MP-SPDZ at fixed Input/Output dimension/size = 8192/67108864. For our implementation, slice size is fixed to 262140. Each subfigure fixes the thread count to the value shown in the caption.}
    \label{fig:party-scaling}
\end{figure*}

We compare our SPDZ implementation on the CPU and GPU back ends against MP-SPDZ while varying the number of parties. We fix the workload to a single linear layer with Input/Output dimension/size = 8192/67108864. For our implementation, slice size is fixed to 262140 when party count is the only variable. To capture interactions between concurrency and multi-party overhead, we report results at three thread settings, $T \in \{1, 8, 32\}$. The resulting performance curves are shown in Figure~\ref{fig:party-scaling}.

Figure~\ref{fig:party-scaling} shows party scaling for all three systems with Input/Output dimension/size = 8192/67108864 and slice size 262140 for our implementation, across $T \in {1, 8, 32}$. In all panels, runtime increases from 2 to 6 parties. Over this range, our CPU back end and MP-SPDZ exhibit roughly linear growth, while our GPU results trend closer to MP-SPDZ at higher party counts when using 8 or more threads.

Again, our CPU back end reports the lowest measured runtime at each party count, followed by our GPU back end, with MP-SPDZ highest for these configurations. 

\subsection{RQ3: Slice Size Scaling}\label{subsec:slice-scaling}

We evaluate how slice size affects performance within our system by comparing the CPU and GPU back ends under fixed workload and party count. Slice size controls the granularity of our batched execution and therefore influences both parallel scheduling and back end efficiency. For MP-SPDZ, its compilation pipeline produces instruction tapes that are executed sequentially within each tape; under multithreading, parallelism is expressed by generating multiple tapes and SIMD within individual instructions.

\begin{figure*}[t]
    \centering
    \small

    \begin{tabular}{ll}
        \hline
        \textbf{Setting} & \textbf{Value} \\
        \hline
        Input/Output dimension / size & 8k/64m \\
        Parties & 2 \\
        Slice size (X-axis) & \{131k, 262k, 524k, 1m\} \\
        \# Computation threads per party & $T \in \{1, 8, 32\}$ \\
        \hline
    \end{tabular}

    \vspace{0.6em}

    \begin{subfigure}[t]{0.32\textwidth}
        \centering
        \includegraphics[width=\linewidth]{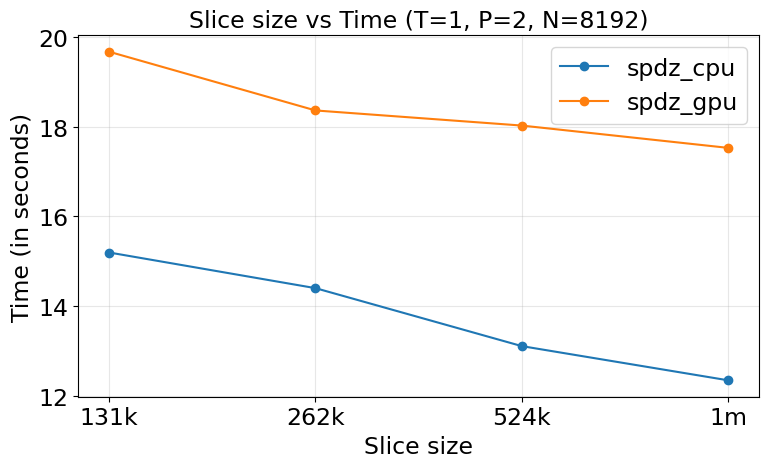}
        \caption{$T=1$}
        \label{fig:slice-scaling-t1}
    \end{subfigure}
    \hfill
    \begin{subfigure}[t]{0.32\textwidth}
        \centering
        \includegraphics[width=\linewidth]{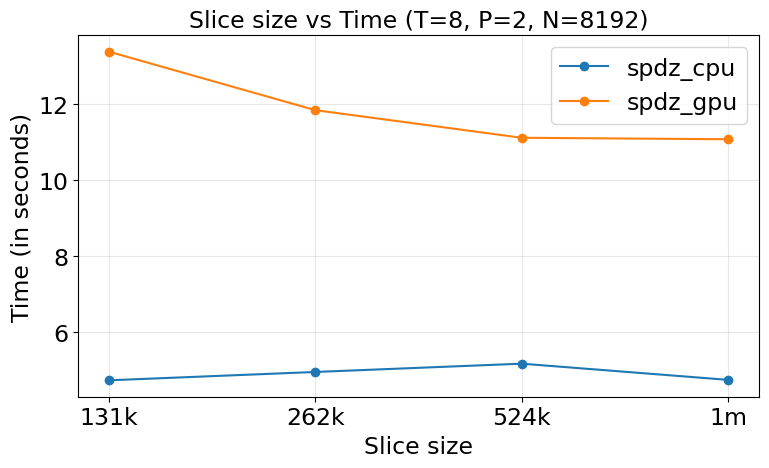}
        \caption{$T=8$}
        \label{fig:slice-scaling-t8}
    \end{subfigure}
    \hfill
    \begin{subfigure}[t]{0.32\textwidth}
        \centering
        \includegraphics[width=\linewidth]{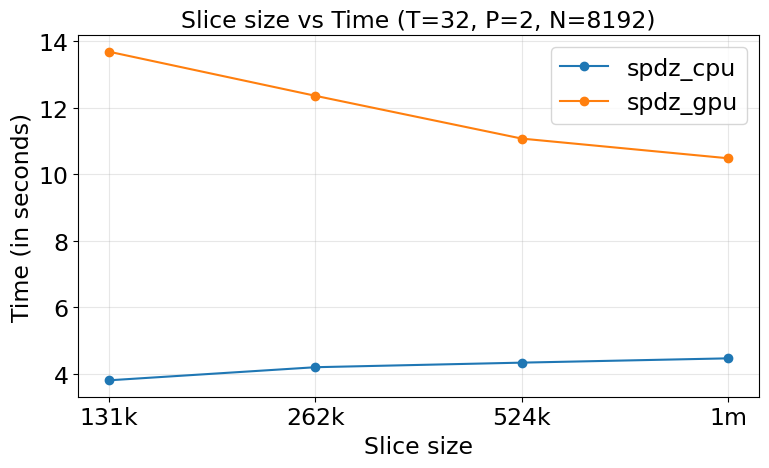}
        \caption{$T=32$}
        \label{fig:slice-scaling-t32}
    \end{subfigure}

    \caption{Slice size sensitivity within our SPDZ implementation at fixed Input/Output dimension/size = 8192/67108864 and 2 parties. Each panel compares our CPU and GPU back ends while varying slice size and holding thread count fixed.}
    \label{fig:slice-scaling}
\end{figure*}

We fix the workload to a single linear layer with Input/Output dimension/size = 8192/67108864 and fix the number of parties to 2. We then vary slice size while holding the thread count fixed, reporting results for $T \in \{1, 8, 32\}$. The results are shown in Figure~\ref{fig:slice-scaling}.

Figure~\ref{fig:slice-scaling} shows that across all three thread settings, the GPU back end reports higher runtime than the CPU back end for the slice sizes evaluated.

At $T=1$, both back ends show a gradual decrease in runtime as slice size increases. The CPU back end decreases from 15.19s at slice 131070 to 12.34s at 1048560, while the GPU back end decreases from 19.67s to 17.53s over the same range. At higher thread counts, the two back ends exhibit different patterns. For $T=8$, CPU times remain within a narrow band (4.73--5.16s) as slice size increases, while the GPU back end decreases from 13.37s to 11.07s. For $T=32$, the CPU back end shows a modest increase from 3.81s to 4.47s across the slice sizes tested, whereas the GPU back end decreases more substantially from 13.68s to 10.48s.

\subsection{RQ4: Input Size Scaling}\label{subsec:input-scaling}

We evaluate how performance scales with input size for our SPDZ implementation on the CPU and GPU back ends and for MP-SPDZ. The workload is a single linear layer, and we vary Input/Output dimension over \{2048, 4096, 8192\}. We fix the number of parties to 2. For our implementation, we additionally fix the slice size to 524280. We report results for three thread settings, $T \in \{1, 8, 32\}$.

\begin{figure*}[t]
    \centering
    \small
    \begin{tabular}{ll}
        \hline
        \textbf{Setting} & \textbf{Value} \\
        \hline
        Input/Output dimension / size (X-axis) & $ \{2k/4m, 4k/16m, 8k/64m\}$ \\
        Parties & 2 \\
        Slice size (ours) & 524280 \\
        \# Computation threads per party & $T \in \{1, 8, 32\}$ \\
        \hline
    \end{tabular}

    \vspace{0.6em}

    \begin{subfigure}[t]{0.32\textwidth}
        \centering
        \includegraphics[width=\linewidth]{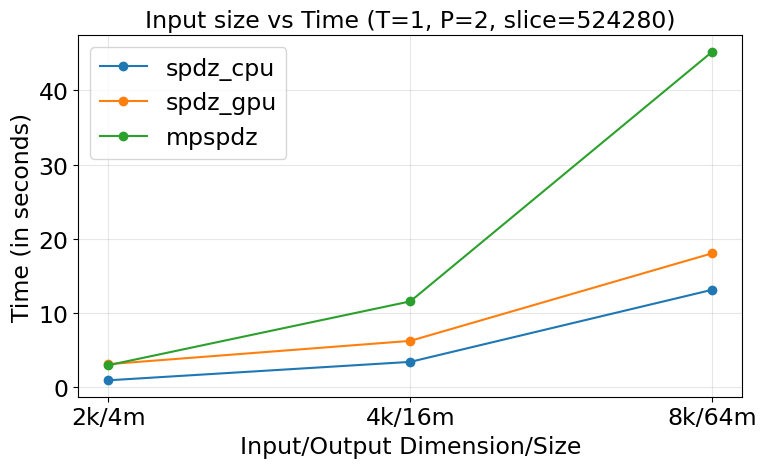}
        \caption{$T=1$}
        \label{fig:input-scaling-t1}
    \end{subfigure}
    \hfill
    \begin{subfigure}[t]{0.32\textwidth}
        \centering
        \includegraphics[width=\linewidth]{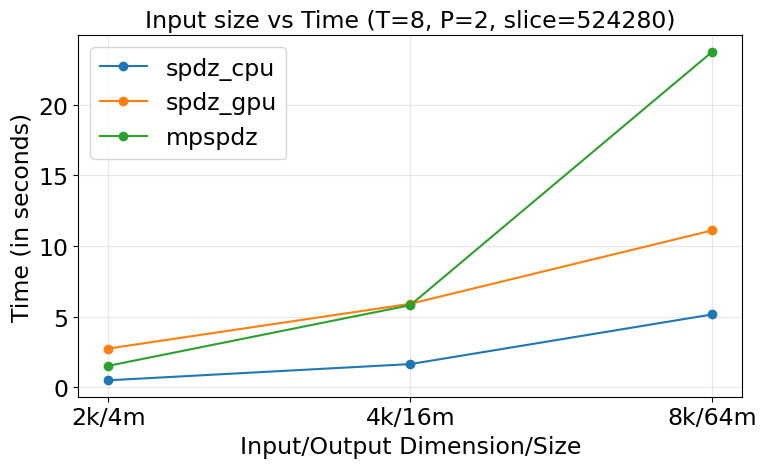}
        \caption{$T=8$}
        \label{fig:input-scaling-t8}
    \end{subfigure}
    \hfill
    \begin{subfigure}[t]{0.32\textwidth}
        \centering
        \includegraphics[width=\linewidth]{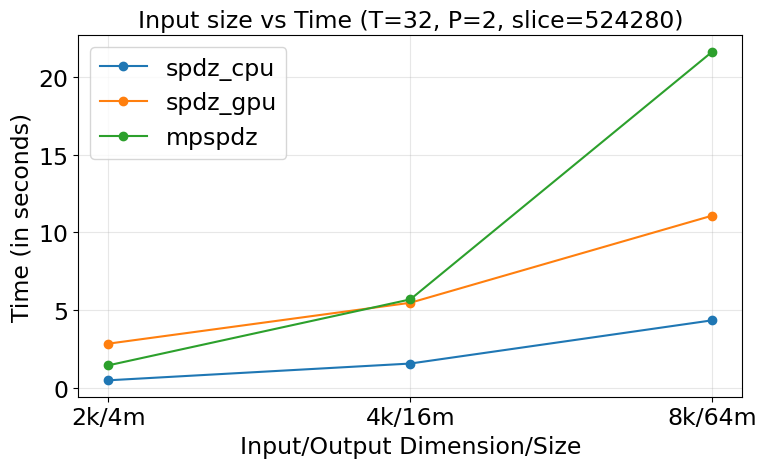}
        \caption{$T=32$}
        \label{fig:input-scaling-t32}
    \end{subfigure}

    \caption{Input size scaling comparison between our SPDZ implementation (CPU/GPU) and MP-SPDZ at 2 parties. For our system, slice size is fixed to 524280. Each panel fixes the thread count and varies Input/Output dimension / size.}
    \label{fig:input-scaling}
\end{figure*}

Figure~\ref{fig:input-scaling} shows consistent input-scaling trends across all three thread settings. Our CPU back end is the fastest configuration at every tested input size. The GPU back end is less favorable at the smallest input but improves its relative position as input size grows. At the largest input (8192), the ordering is consistent across all thread settings: our CPU back end is lowest, our GPU backend is next, and MP-SPDZ is highest.

The observed growth factors align with the quadratic structure of the workload: when doubling the input dimension from 2048 to 4096 and from 4096 to 8192, MP-SPDZ scales by roughly $\approx 3.80$--$4.08\times$ per step across all three thread settings. Our CPU back end shows comparable but slightly smaller per-step increases (about $2.78$--$3.88\times$). In contrast, our GPU back end exhibits the smallest multiplicative increases (about $1.88$--$2.90\times$), suggesting better scaling for larger linear layers within the tested range.

\subsubsection{Relative performance vs.\ MP-SPDZ}
Across all thread settings, our CPU back end outperforms MP-SPDZ by a wide margin, ranging from about $2.98$--$3.28\times$ faster at $N=2048$ to about $4.60$--$4.97\times$ faster at $N=8192$ for $T \in \{8,32\}$. Equivalently, our CPU runtime is roughly 66--70\% lower than MP-SPDZ at $N=2048$, and this advantage grows with input size and threads, reaching about 79.9\% lower runtime at $N=8192$ with $T=32$. 

Our GPU back end shows a different pattern: at $N=2048$ it is slower than MP-SPDZ (about $1.05\times$ slower at $T=1$ and $1.80$--$1.96\times$ slower at $T=8$ and $T=32$), but the gap narrows quickly as input size increases. At $N=4096$, the GPU back end is competitive with MP-SPDZ, ranging from $1.86\times$ faster at $T=1$ to near parity at $T=8$ and $T=32$. At $N=8192$, the GPU back end consistently surpasses MP-SPDZ by about $1.95$--$2.51\times$ across $T\in\{1,8,32\}$ (roughly 49--60\% lower runtime). This crossover behavior, together with the smaller per-doubling growth factors, indicates that our GPU back end scales well over the tested input range and benefits more from increased problem size than MP-SPDZ.

\subsection{RQ5: Runtime Component Breakdown}\label{subsec:runtime-component}

We next report a coarse-grained runtime component breakdown for our SPDZ implementation and MP-SPDZ under the same linear-layer microbenchmark.

We group runtimes into three categories: \emph{front end}(MPC program compilation), \emph{setup}(fetch necessary data, establish socket connection across parties, etc.), and \emph{online}(input sharing \& computation).

\begin{figure}[t]
    \centering
    \small
    \begin{tabular}{ll}
        \hline
        \textbf{Setting} & \textbf{Value} \\
        \hline
        Input/Output dimension / size & 8k/64m \\
        Parties & 2 \\
        \# Computation threads per party & 8 \\
        Slice size (ours) & 524k \\
        Stages & front end, setup, online \\
        \hline
    \end{tabular}

    \vspace{0.6em}

    \includegraphics[width=0.95\linewidth]{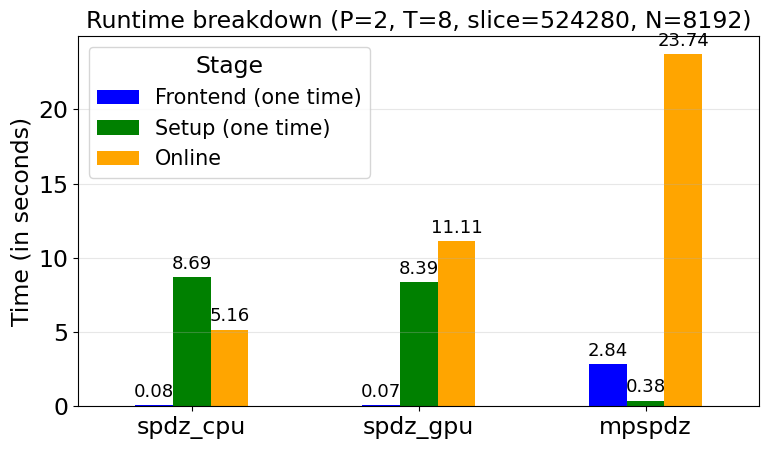}
    \caption{Runtime breakdown for our SPDZ implementation (CPU/GPU) and MP-SPDZ on a fixed linear layer. Front end is a one time cost per program; setup is a one time cost per input run.}
    \label{fig:runtime-breakdown}
\end{figure}

We fix the number of parties to 2, thread count to 8, and evaluate the same linear-layer workload with Input/Output dimension/size = 8192/67108864. For our implementation, we fix the slice size to 524280. Figure~\ref{fig:runtime-breakdown} summarizes the resulting stage-level runtimes.

The figure shows that for our implementation, front end time is negligible on both CPU and GPU, and the total runtime is largely driven by setup and online execution. At this setting, setup is the largest single component for our CPU back end and remains substantial for our GPU back end. In contrast, MP-SPDZ shows a noticeably larger front end cost but a much smaller setup stage, and its total time is dominated by the online phase. The online component is smallest for our CPU back end, larger for our GPU back end, and largest for MP-SPDZ under the same workload and thread setting across the three systems.

\subsection{Usability}\label{subsec:usability}

Our system does not require a domain-specific language. Instead, the front end is a restricted subset of C with lightweight privacy annotations that mark values as private or public. This design keeps the programming model close to standard systems code and lowers the barrier to adopting our toolchain for existing workloads.

The adaptation can be even broader. The compiler pipeline is built around LLVM IR. Therefore, with a reasonable amount of front end plumbing, any language that can compile to LLVM IR can serve as an input language for our system. This provides a path to supporting multiple high-level languages without changing the core optimization and lowering pipeline, and enables programmers to work in familiar ecosystems while benefiting from our back end and runtime.

\section{Discussion}\label{sec:discussion}

Our results demonstrated solid improvement in performance for compute-heavy SPDZ workloads. At the same time, they highlight several practical limitations that clarify when these benefits are expected to be largest. We summarize these limitations and outline our future plans.
\subsection{Workload Dependence and Limited Parallelism}
Our design aims to overlap computation with other computation and with communication. This strategy is most effective when the computation graph exposes sufficient parallelism and when batching can amortize protocol and scheduling overheads. Consequently, our approach may offer limited advantages under two conditions.

First, for workloads that form long dependent chains (i.e., deep computation trees with few independent nodes), there is inherently less opportunity for parallel execution. In this regime, dependency management and scheduling overheads can dominate, reducing the potential gains from our parallelization strategy.

Second, when computations are not sufficiently compute-heavy or cannot be batched effectively, communication costs may dominate runtime. Since our system is primarily designed for algebra-heavy settings, these communication-bound workloads fall outside the intended operating regime of our design.

These limitations are expected. Our system intentionally targets large, parallelizable, compute-heavy workloads and is designed to surface and exploit their structural parallelism through aggressive batching and scheduling. In contrast, more general MPC frameworks aim to be broadly versatile across workloads and deployment settings, which can limit the extent to which they fully capitalize on the parallelism available in these specialized, algebra-dense computations.

\subsection{Performance and Deployment Assumptions}\label{gpu_performance}
\paragraph{GPU Performance}{Our current GPU back end underperforms the CPU in several settings. A key reason is that our evaluation places all parties (implemented as separate OS processes) on the same physical GPU or thread pools. As the number of parties and threads per party increases, this shared-device configuration introduces contention, serialization, and queueing overheads that can outweigh the accelerator's raw throughput advantage. This effect is consistent with our observed trend that GPU performance degrades with increasing party count, particularly under larger thread settings (see Figure ~\ref{fig:party-scaling}).}

\paragraph{Hyper-threading}{The flattening of performance beyond 16 threads per party across all three systems is largely a consequence of CPU oversubscription on our evaluation machine. We allocate 64 CPU cores per experiment; as we increase both the number of parties and the threads per party, the total number of worker threads approaches and then exceeds the number of physical cores. Beyond this point, additional threads primarily rely on hyper-threading rather than new physical parallelism, leading to the observed flattening trend in performance.}

\paragraph{Future work}
We will evaluate more realistic deployment models where each party has its own GPU resources. On the CPU side, we will repeat our scaling experiments on machines with more physical cores.

\subsection{Setup-Time Bottlenecks and I/O Overheads}
Although our design focuses on online performance, our current implementation shows higher setup time than MP-SPDZ. This is because our setup stage is deliberately simple, single-threaded, and unoptimized: inputs, preprocessed data, and MPC graphs are stored in JSON and loaded through a generic serialization layer, incurring substantial parsing and I/O overhead.

\paragraph{Future work}
We plan to replace the JSON-based input pipeline with a more efficient representation, such as fixed-width binary encodings, and to reorganize preprocessed data into cache-friendly layouts. These changes should significantly reduce setup time and alleviate the current setup bottleneck, allowing the improvements in online performance to translate more directly into end-to-end speedups.

\subsection{Configuration Sensitivity and Autotuning}
Performance depends on several interacting parameters, including CPU thread count, slice size, and vectorization settings. While batching is automated for directly compiled C programs via LLVM and handled by our runtime scheduler for black-box kernels such as the linear layer, selecting the best slice or batch size can still require manual tuning. In particular, the optimal slice size may depend on thread count, which neither LLVM nor our current analyzer models explicitly.

Importantly, even without optimization-aware tuning, users still benefit from much of our parallelization and scheduling logic, preserving usability while allowing experts to extract additional performance when needed.

\paragraph{Future work.}
We plan to introduce autotuning and profile-guided parameter selection, potentially via a lightweight calibration step that improves portability across deployment settings.

\subsection{Usability and Static Safety Checks}
To enforce SPDZ's security requirements, our current implementation rejects programs whose control flow depends on secret values by failing loudly at runtime. This approach ensures no secret leakage, but it can waste time and compute resources by surfacing violations late in the development cycle.

\paragraph{Future work.}
We plan to move these checks to compile time by incorporating secret-dependent control-flow analysis into the front end. We also aim to refine and streamline the privacy annotation interface to reduce user burden and improve readability.

\section{Related Work}\label{sec:related-work}

Secure multiparty computation (MPC) has advanced from foundational protocols to general-purpose, performance-oriented systems. Early milestones include Yao’s garbled circuits for two-party computation, the GMW generalization to the multiparty setting, and BGW’s information-theoretic honest-majority construction via Shamir secret sharing \cite{yao_cite,gmw_cite,bgw_cite,secret_sharing_gen_cite}. Practicality improved with OT extension and early secure-function-evaluation systems such as Fairplay \cite{ot_extension_cite,fairplay_cite}. The 2010s brought actively secure, dishonest-majority arithmetic MPC into wider use through the SPDZ family and MASCOT’s OT-based triple generation, followed by ring-based and ML-oriented variants such as SPDZ2k and ABY3 \cite{spdz_cite,spdz2_practical_cite,mascot_cite,CramerDESX18SPDZ2k,aby3_cite}. Recent work improves both cryptographic subroutines and systems performance, including batching, scheduling, and accelerator support \cite{overdrive_cite,cryptgpu_cite,crypten_ml_cite}.

MP-SPDZ is a prominent protocol-rich framework that unifies many protocols behind a custom virtual machine and compiler \cite{mpspdz_cite}. This enables convenient cross-protocol evaluation, but requires applications to target its VM and high-level front ends rather than reusing existing compiler infrastructures such as LLVM. A lower-level C++ interface exists, but it operates close to protocol objects and networking primitives and is mainly intended for advanced use.

PICCO similarly aims to generalize secure computation across applications\cite{picco_cite}, but its primary contribution is at the front end, compiling an extended C-like language via source-to-source translation into MPC code. In contrast, we focus on the compiler middle end, operating on LLVM IR to leverage and extend existing transformation and optimization passes.

A separate line of work introduces MPC-specific intermediate representations. FUSE defines a custom IR and file format for circuit-level optimizations before lowering to backends \cite{fuse_cite}. COMBINE and related systems also adopt a custom SSA-style MPC IR and argue that directly reusing generic compiler IRs is nontrivial, particularly around representing data-oblivious control flow and $\phi$-merges in MPC \cite{combine_cite}. These IRs further incorporate array- and SIMD-aware constructs to support backend-independent vectorization and scheduling.

Sequre introduces a Python-like DSL that compiles through a custom intermediate form before lowering to LLVM and native code \cite{sequre_cite}. It emphasizes cryptographic efficiency through compile-time rewrites and analyses that reduce expensive MPC operations and select efficient instantiations. Its evaluation primarily targets CPU execution and places less emphasis on multi-threaded scheduling or accelerator backends.

Our work occupies a different point in this design space. Rather than introducing a new VM or MPC-specific IR, we keep LLVM IR as the central representation and build MPC-aware analyses and transformations around it. We restrict the accepted fragment of C so that each surviving IR instruction has clear MPC semantics, recover MPC-relevant structure from dataflow and control-flow, and reuse LLVM’s vector types and passes to expose batched arithmetic. We then map these batched instructions to MPC-specific operations and execute them on pluggable SIMD CPU and GPU back ends. 

\section{Conclusion}\label{sec:conclusion}
We presented a proof-of-concept LLVM-based optimization pipeline for the SPDZ protocol that leverages a mainstream C front end, mature compiler analyses, and a runtime scheduler to reduce the burden of manually expressing parallelism. By automatically batching independent arithmetic operations and overlapping computation with communication, our system improves online performance on controlled microbenchmarks. Our evaluation against MP-SPDZ shows that the CPU back end achieves up to $5.56\times$ speedup under intermediate and heavy algebraic workloads and scales well with thread count, while the GPU back end benefits from larger input sizes. These results suggest that integrating LLVM with protocol-specific scheduling is a promising architectural direction for improving the performance and usability of actively secure arithmetic MPC, and offers a complementary path for future MPC framework designs.

\section*{Acknowledgments}\label{sec:acknowledgments}

We thank the support from the Davidson Research Infrastructure team and, in particular Michael Blackmon and Umut Turk. We also thank Davidson College and the Davidson Research Initiative program for their general Summer support.


\bibliographystyle{IEEEtran}
\bibliography{references}

\end{document}